# New bulk p-type skutterudites DD$_{0.7}$Fe$_{2.7}$Co$_{1.3}$Sb$_{12-x}$X$_x$ (X = Ge, Sn) reaching ZT>1.3

G. Rogl$^{a,b,d,e,*}$, A. Grytsiv$^{a,b,e}$, P. Heinrich$^b$, E. Bauer$^{b,e}$, P. Kumar$^c$,
N. Peranio$^c$, O. Eibl$^c$, J. Horky$^d$, M. Zehetbauer$^d$, P. Rogl$^{a,e}$

$^a$Institute of Physical Chemistry, University of Vienna, Währingerstrasse 42, A-1090 Wien, Austria
$^b$Institute of Solid State Physics, Vienna University of Technology, Wiedner Hauptstrasse 8-10, A-1040 Wien, Austria
$^c$Eberhard Karls University of Tübingen, Institute of Applied Physics, Auf der Morgenstelle 10, 72076 Tübingen, Germany
$^d$Research Group Physics of Nanostructured Materials, University of Vienna, Boltzmanngasse 5, A-1090 Wien, Austria
$^e$Christian Doppler Laboratory for Thermoelectricity, Wien, Austria

**Abstract**

The best p-type skutterudites so far are didymium filled, Fe/Co substituted, Sb-based skutterudites. Substitution at the Sb-sites influences the electronic structure, deforms the Sb$_4$-rings, enhances the scattering of phonons on electrons and impurities and this way reduces the lattice thermal conductivity.

In this paper we study structural and transport properties of p-type skutterudites with the nominal composition DD$_{0.7}$Fe$_{2.7}$Co$_{1.3}$Sb$_{11.7}$\{Ge/Sn\}$_{0.3}$, which were prepared by a rather fast reaction-annealing-melting technique. The Ge-doped sample showed impurities, which did not anneal out completely and even with ZT > 1 the result was not satisfying. However, the single-phase Sn-doped sample, DD$_{0.7}$Fe$_{2.7}$Co$_{1.3}$Sb$_{11.8}$Sn$_{0.2}$, showed a lower thermal and lattice thermal conductivity than the undoped skutterudite leading to a higher ZT=1.3, hitherto the highest ZT for a p-type skutterudite. Annealing at 570 K for 3 days proved the stability of the microstructure. After severe plastic deformation (SPD), due to additionally introduced defects, an enhancement of the electrical resistivity was compensated by a significantly lower thermal conductivity and the net effect led to a record high figure of merit: ZT = 1.45 at 850 K for DD$_{0.7}$Fe$_{2.7}$Co$_{1.3}$Sb$_{11.8}$Sn$_{0.2}$.

1. **Introduction**

Thermoelectricity, as an environment-friendly and reliable technology, is attracting growing attention in the field of power generation as well as for waste heat recovery [1-3]. Filled skutterudites are the most potential thermoelectric (TE) materials in temperature gradients of 300 to 850 K because they not only have excellent TE properties but also show appropriate mechanical properties, a necessary requisite for TE device engineering [4-11]. The efficiency of a TE material is determined by its dimensionless figure of merit $ZT = S^2T/\rho\lambda$, where S is the Seebeck coefficient (or thermopower), $\rho$ is the electrical resistivity, $\lambda$ is the thermal conductivity ($\lambda = \lambda_e + \lambda_{ph}$), consisting of the electronic thermal conductivity $\lambda_e$ and lattice thermal conductivity $\lambda_{ph}$, and T is the temperature. A good TE material has a high power factor $pf = S^2\sigma$ ($\sigma$ is the electrical conductivity with $\sigma = 1/\rho$) combined with a low thermal conductivity [12,13]. For thermoelectric power generation not only a high ZT value is necessary, but a high average ZT ($ZT_a$) over the temperature regime exploited. The so-called thermal-electric conversion efficiency $\eta$ (in %) is calculated according to equation (1)

$$\eta = \frac{T_h - T_c}{T_h} \frac{\sqrt{1+(ZT_a)} - 1}{\sqrt{1-(ZT_a)} + \frac{T_c}{T_h}} \tag{1}$$

including the Carnot efficiency (where $T_h$ and $T_c$ are the temperatures on the hot and cold side respectively) as well as ($ZT_a$).

Filled skutterudites, MT$_4$X$_{12}$ (M is a rare earth, an actinide, an alkaline-earth, or an alkaline metal, T is a transition metal of subgroup VIII, and X is a pnictogen atom), are compounds with a body centered cubic crystal structure (space group $Im\bar{3}$ [14]) in which M atoms reside in the large icosahedral voids formed by tilted octahedra. The physics of filled skutterudites is governed by the interplay of filler ions and their host structure [15-21]. For an excellent TE performance not only the optimal filler elements and filling fraction need to be designed but also single-element or multi-element doping as well as the combination of filling and doping have led to high ZT values [1,12,13,19, 22-61]. The highest ZT values for both p- and n-type skutterudites have been renewed continuously, recently reaching ZT = 1.2 for double or multifilled p-type and ZT = 1.7 for multi filled n-type skutterudites [49,50,54,60,62-64]. These high values could be topped applying severe plastic deformation (SPD) via high-pressure torsion (HPT) yielding ZT ~ 1.3 and ZT ~ 1.9 respectively [64-70]. However, achieving comparable ZT values for p- and n-type skutterudites is crucial for improving the conversion efficiency of TE devices, demanding further improvement for p-type skutterudites.

Among p-type skutterudites, the highest ZT values (~1.2) could be gained for DD$_y$(Fe$_{1-x}$Co$_x$)$_4$Sb$_{12}$, with didymium (DD) as natural double filler at a filling level y=0.60±0.1 and for x = 0.25±0.05 [49,50,60]. The reasons for these high ZT values are threefold: (i) alloy phonon scattering as DD consists of Pr (4.76 mass%) and Nd (95.24 mass%); (ii) the large atomic mass of DD in respect to a small ionic radius leads to low frequency modes further reducing the thermal conductivity; and (iii) the introduction of structural disorder via T-metal substitution (Fe/Co) enhances phonon scattering and lowers the lattice thermal conductivity. It could be shown in several studies that a substitution at the Sb-sites influences the electronic structure, deforms the $Sb_4$-rings [70] and introduces defects, all in favor of a low lattice thermal conductivity [61,71-78].

The production of a bulk nanostructured material should be simple, economic and fast. One way, optimized and so far very successfully, is to ball mill (BM) and hot press (HP) the material after a melting/reaction step [43,46,49,50,60,62,64,70,78]. In addition SPD via HPT was used to further improve the TE quality of the sample as has been shown by the authors successfully for other filled and unfilled skutterudites [55,65-70,78,79]. HPT introduces deformation-induced defects, mostly dislocations and vacancies, enhances the formation of grain boundaries and reduces the crystallite size, resulting in a substantial reduction of the lattice thermal conductivity via phonon scattering [55,65-70,78,79,80-85].

In this work we studied the influence of doping Sb by Ge or Sn on p-type skutterudites with the nominal composition, $DD_{0.7}Fe_{2.7}Co_{1.3}Sb_{11.7}\{Ge,Sn\}_{0.3}$. Several tasks were pursued: (i) physical properties, especially thermal conductivity, were compared with the physical properties of undoped $DD_{0.60}Fe_{2.8}Co_{1.2}Sb_{12}$ of a previous work [50]; (ii) thermal sample stability was investigated by various annealing programs; (iii) the change of the physical properties by severe plastic deformation via high pressure torsion was studied in respect to defects introduced, smaller crystallite size and an enhancement of grain boundaries.

**2. Experimental**

Two samples with the overall formula $DD_{0.7}Fe_{2.7}Co_{1.3}Sb_{11.7}X_{0.3}$ (X = Ge, Sn) were prepared from master alloys $Fe_{2.7}Co_{1.3}Sb_{11.7}(Ge/Sn)_{0.3}$ via an optimised melting reaction technique from stoichiometric amounts of high purity elements (Fe, 99.5%, wire, Co, 99.9%, powder < 150 micron, Sb, 99.95%, crystals, Ge, 99.999 pieces, and Sn, 99.9 shot) by mixing, sealing into evacuated quartz tubes, melting at 950°C followed by air quenching (details are given in refs. [46,49,50,86,87]). Then the stoichiometric amount of DD (from Treibacher Industrie AG, 99.9%) was added. The samples were then sealed under vacuum into quartz tubes, quickly heated to 600°C, then slowly (2°C/min) heated to 720°C and after a fast rise (5°/min) of the temperature melted at 950°C and air quenched. The reguli were ground in the glove box to obtain particles < 50 μm, followed by ball milling in Ar-filled tungsten carbide vessels (Fritsch planetary mill Pulverisette 4). Both samples were hot-pressed (FCT uniaxial hot-press HP W 200/250 2200-200-KS, Ar, 700°C, 56 MPa, 30 min).

X-ray powder diffraction data were collected with monochromated $CuK_{\alpha_1}$- and $FeK_{\alpha_1}$-radiation with a HUBER-Guinier image plate recording system using 99.9999% pure Si as internal standard. X-ray spectra were used to calculate the lattice parameters (program STRUKTUR [88]) and the program FULLPROF was employed [89] for quantitative Rietveld refinement in order to determine the total filling level in combination with detailed chemical composition analyses by electron probe microanalysis (EPMA – EDX) with an INCA Penta FETx3 – Zeiss SUPRA$^{TM}$55VP equipment.

The crystallite size and the residual strain were evaluated from the XPD patterns (spectra from the $FeK_{\alpha_1}$-radiation) using the MDI JADE 6.0 software (Materials Data Inc., Liverpool, CA). With this method the crystallite size was calculated from the full width at half-maximum (FWHM) of a single diffraction peak using the formula developed by Scherrer [90]. Si was used as internal standard to account for instrumental broadening. Three well separated reflections, (240), (332) and (422), within a $2\theta$ range from 1° to 58° of the X-ray profile were used. For details see ref. [79,86].

Ge-doped samples were analyzed by analytical transmission electron microscopy (TEM) with a TEM Zeiss 912 Omega, operated at 120 kV and equipped with an omega energy filter and an energy-dispersive X-ray (EDX) detector. Energy-filtered bright-field and dark-field images were acquired for imaging grains and dislocations, using two-beam conditions for diffraction contrast-based imaging. Selected area electron diffraction patterns were achieved using aperture-selected areas for diffraction of 750 nm in size. A superposition of energy-filtered images (RGB) acquired at 17 eV (red), 31 eV (green), and 17 eV (blue) turned out to be most suitable for imaging secondary phases, rather than 17 eV (red), 31 eV (green), and 59 eV (blue) or other RGB combinations of these energy filtered images. Point EDX spectra were acquired with a spot size of 32 nm. Details of the quantitative chemical analysis by EDX were explained elsewhere [66].

Electrical resistivity and Seebeck coefficient (room temperature to 550°C) were measured simultaneously using an ULVAC-ZEM3 (Riko, Japan) equipment. The thermal conductivity above room temperature was calculated from the thermal diffusivity $D_t$ measured by a flash method (Flashline-3000, ANTER, USA), specific heat $C_p$ and density $d_m$ using the relationship $\lambda = D_t C_p d_m$. Measurement errors for the electrical resistivity and Seebeck measurements are < 3%, for thermal conductivity ~ 5% (leading to an error for ZT<12%), which all are in the range confirmed from parallel measurements in other laboratories (see Ref. [87]; ρ and S data were also confirmed with a LSR-3 measurement system (LINSEIS, Germany) in the Indian Institute of Science, Dept. of Physics in Bangalore). Measurements of the HPT processed samples with increasing and decreasing temperature were all in the same range; data after measurement induced heating showed stability.

Electrical resistivity and Hall data from 4 K to room temperature were obtained by a physical properties measurement system (PPMS) from Quantum Design with a standard six-point method in a magnetic field up to 9 T.

The density $d_m$ in g/cm$^3$ of both samples was obtained by the Archimedes' method, using distilled water. The relative densities $d_{rel}$ (in %) were calculated, using the X-ray density $d_X = (MZ)/(VN)$ where M is the molar mass, Z is the number of formula units per cell, N is Loschmidt`s number and V is the volume of the unit cell.

Both samples were HPT processed with 4 GPa and one revolution at room temperature (shear strain of about 30) with an equipment from W. Klement, Austria; for all details of HPT processing of skutterudites the reader is referred to refs. [65-69,79].

**3. Results and Discussion**

**3.1. Structural properties**

X-ray intensity patterns of both skutterudites with the nominal composition $DD_{0.7}Fe_{2.7}Co_{1.3}Sb_{11.7}Ge_{0.3}$ and $DD_{0.7}Fe_{2.7}Co_{1.3}Sb_{11.7}Sn_{0.3}$ were completely indexed on the base of a body-centred cubic lattice prompting isotypism with the ordered $LaFe_4P_{12}$-type. As the traces of secondary phases were negligible, one can state that a pure skutterudite phase was observed for the Sn-doped sample (Fig.1) whereas secondary phases were detected for $DD_{0.7}Fe_{2.7}Co_{1.3}Sb_{11.7}Ge_{0.3}$ besides the dominant skutterudite phase. Microstructures were analyzed by combined SEM and EPMA measurements. The Sn-doped sample showed grains in nano size homogeneously distributed; no other phases than the skutterudite phase was observed. For the Ge-doped sample secondary phases could be detected in small amounts (less than 2 vol% $Fe_3Ge_2Sb$ and even smaller amounts of $NdSb_2$). Zhang et al. [61] found $Fe_3Ge_2Sb$ as secondary phase in $Nd_{0.6}Fe_2Co_2Sb_{12-x}Ge_x$ skutterudites for samples with higher Ge content (x > 0.5), whereas Yu et al. [91] reported no secondary phases for Ge-doped skutterudites $Ba_{0.3}In_{0.2}FeCo_3Sb_{12-x}Ge_x$ (x = 0 to x = 0.4). Because of the strong correlation with the temperature factor $B_{iso}$, the occupancies of Fe/Co in the 8$c$ site and the occupancies for Sb and Ge as well as for Sb and Sn in 24$g$ were introduced from EPMA data. Afterwards the occupancy of DD in the 2$a$ position was refined, and finally all occupancies were fixed in order to refine the $B_{iso}$-values for these atom sites. DD contents determined from this procedure agree well with the data obtained from EPMA; nominal and actual compositions ($DD_{0.54}Fe_{2.7}Co_{1.3}Sb_{11.9}Ge_{0.1}$ and $D_{0.59}Fe_{2.7}Co_{1.3}Sb_{11.8}Sn_{0.2}$) are listed in Tab.1 together with the lattice parameters before and after annealing, after HPT processing and after measurement induced annealing. It should be noted, that the actual DD-levels in both skutterudite phases are only slightly smaller in comparison with the nominal content. Similarly the Sn-content is marginally reduced, but the Ge-content dropped significantly in accordance with the formation of secondary phases containing a higher level of Ge.

Table 1. Nominal composition, actual composition (from combined EPMA-Rietveld refinement), lattice parameter, a, in nm, average crystallite size, $s_c$, in nm, grain size, $s_g$, in μm, relative density, $d_{rel}$ in %

| Nominal composition | Actual composition | | a | $s_c$ | $s_g$ | $d_{rel}$ |
|---|---|---|---|---|---|---|
| reference sample | $DD_{0.60}Fe_{2.8}Co_{1.2}Sb_{12}$ | HP | 0.91104(3) | - | - | 98.0 |
| $DD_{0.7}Fe_{2.7}Co_{1.3}Sb_{11.7}Ge_{0.3}$ | | powder | 0.91042(6) | 210(8) | - | - |
| | $DD_{0.54}Fe_{2.7}Co_{1.3}Sb_{11.9}Ge_{0.1}$ | HP | 0.91037(4) | 330(10) | 1-2 | 99.1 |
| after 3 days at 570°C | | HT1 | 0.91058(1) | 340(10) | 1-2 | 98.5 |
| after HPT | | HPT | 0.91051(3) | 100(5) | 0.02-0.4 | 94.0 |
| after 5h from RT to 550°C | | HPT+HT2 | 0.91047(1) | 254(7) | 0.04-0.6 | 95.8 |
| $DD_{0.7}Fe_{2.7}Co_{1.3}Sb_{11.7}Sn_{0.3}$ | | powder | 0.91061(2) | 200(5) | - | - |
| | $DD_{0.59}Fe_{2.7}Co_{1.3}Sb_{11.8}Sn_{0.2}$ | HP | 0.91033(2) | 336(10) | 1-2 | 99.1 |
| after 3 days at 570°C | | HT1 | 0.91035(3) | 335(9) | 1-2 | 98.9 |
| after HPT | | HPT | 0.91046(4) | 95(3) | 0.02-0.4 | 93.8 |
| after 5h from RT to 550°C | | HPT+HT2 | 0.91039(3) | 256(8) | 0.04-0.6 | 95.5 |

The lattice parameter of $DD_{0.59}Fe_{2.7}Co_{1.3}Sb_{11.8}Sn_{0.2}$ (Tab.1, Fig.2a, A) as well as of $DD_{0.54}Fe_{2.7}Co_{1.3}Sb_{11.9}Ge_{0.1}$ (Tab. 1, Fig. 2b, A) became slightly smaller after compacting the powder in the hot press. But whereas the lattice parameter of the Ge-doped sample became slightly larger after annealing at 570°C for 3 days, the Sn-doped sample showed no change. In parallel, also the density (Tab. 1, Figs. 2a and 2b, B) and transport properties of $DD_{0.59}Fe_{2.7}Co_{1.3}Sb_{11.8}Sn_{0.2}$ were not affected by the heat treatment. Thus it was concluded that the single-phase Sn-sample did not undergo any structural changes during the heat treatment, whereas the Ge-doped sample obviously did undergo changes. The X-ray pattern and EPMA investigations confirmed that both secondary phases, $Fe_3Ge_2Sb$ and $NdSb_2$ of $DD_{0.54}Fe_{2.7}Co_{1.3}Sb_{11.9}Ge_{0.1}$ vanished almost completely after annealing, but there was no change in the composition of the main skutterudite phase.

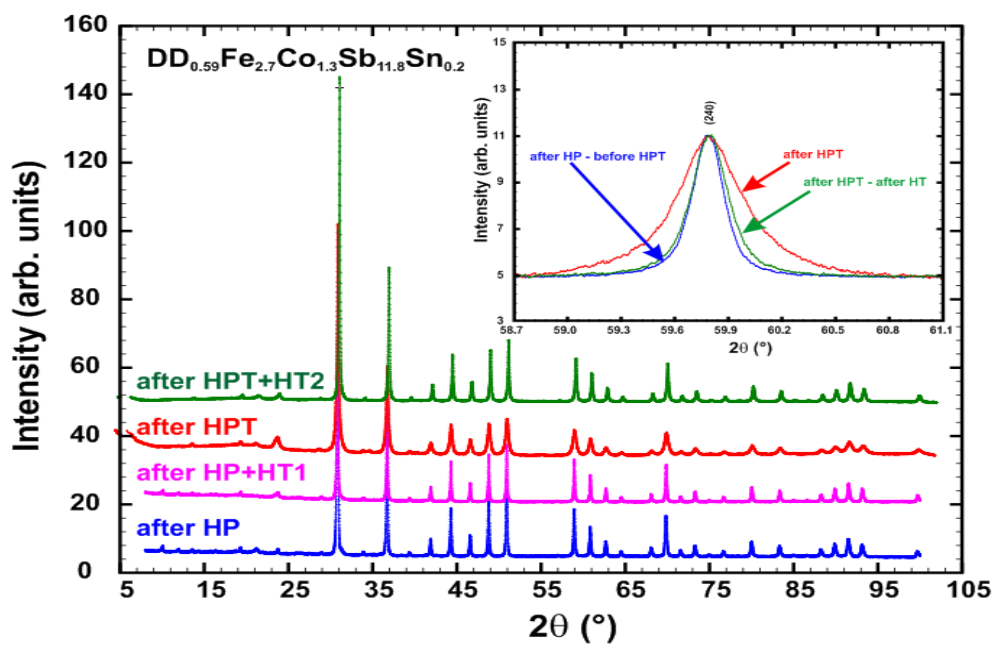

Fig. 1. X-ray intensity pattern of $D_{0.59}Fe_{2.7}Co_{1.3}Sb_{11.8}Sn_{0.2}$ after HP, after HP + HT1, after HPT and after annealing the HPT sample (HPT + HT2). Insert: (240) peak after HP, after HPT and after annealing the HPT sample.

Fig. 1 displays X-ray patterns of $DD_{0.59}Fe_{2.7}Co_{1.3}Sb_{11.8}Sn_{0.2}$, normalized to the highest skutterudite peak, right after HP, after annealing at 570°C for 3 days (heat treatment 1, HT1), after HPT processing and after measurement induced heating (HT2) of the HPT sample. All X-ray intensity patterns show negligible traces of secondary phases such as for instance $FeSb_2$. Comparing the peak widths of the X-ray spectra, no change after annealing of the HP sample (HP + HT1) can be seen, but broadening after HPT processing indicates at least smaller crystallite sizes. After measurement induced heating the peak widths are slimmer again, indicating an additional change in the crystallite size. After measurement induced heating the peak widths are slimmer again, indicating an additional change in the grain size. These changes of the peak widths (Tab. 1) are better visible in the insert of Fig. 1 (the peak "after HP and HT1" was not plotted because it almost exactly matches the peak "after HP").

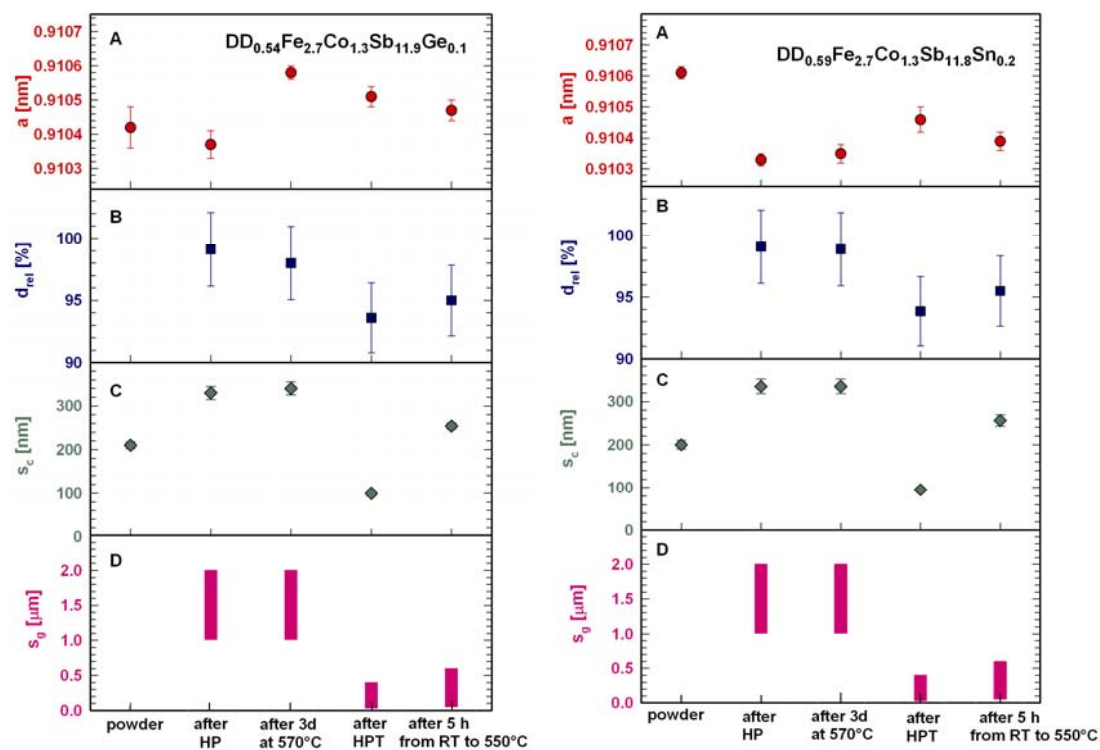

Fig. 2 a: A: lattice parameter, a, B: relative density, $d_{rel}$, C: average crystallite size, $s_c$, D: grain size, $s_g$, of $D_{0.59}Fe_{2.7}Co_{1.3}Sb_{11.8}Sn_{0.2}$ as powder, HP, after HPT and after respective heat treatments.

Fig. 2 b: A: lattice parameter, a, B: relative density, $d_{rel}$, C: average crystallite size, $s_c$, D: grain size, $s_g$, of $D_{0.54}Fe_{2.7}Co_{1.3}Sb_{11.9}Ge_{0.1}$ as powder, HP, after HPT and after respective heat treatments.

Quantitative analyses reveal that for both samples the crystallite size, $s_c$, of the powder (Tab. 1, Figs. 2a and 2b, C) right after ball-milling (powder) is enhanced after hot pressing (after HP) from about an average $s_c = 200$ nm to an average $s_c = 335$ nm, but there is no further change after the additional annealing process (after 3 days at 570°). After severe plastic deformation by HPT, however, the crystallite size is reduced to $s_c \sim 95$ nm, which is less than one third of the original size but it is "growing" back to $s_c \sim 260$ nm (after 5 h from RT to 550°C) after annealing the HPT processed sample. The residual strain in the HPT processed sample is 2.5 times higher than before.

The grain size, $s_g$, (Tab. 1, Figs. 2a and 2b, D) determined from broken surface images (SEM) as well as from TEM investigations, after HP is in the range of 1-2 μm average size with no change after annealing, but it is reduced to 0.02 - 0.4 μm after HPT and grows back to 0.04 - 0.6 μm after annealing the HPT sample.

Not only the crystallite size and strain of $D_{0.59}Fe_{2.7}Co_{1.3}Sb_{11.8}Sn_{0.2}$ changed after HPT and measurement induced heating, but also the lattice parameter and density (see Tab.1 and Figs. 2a and 2b, A, B). After HPT processing the lattice parameter is enlarged, due to vacancies and/or very fine micro cracks and accordingly the density is lower. These changes were also found for other skutterudites and clathrates after SPD [65-69,79,92]. It is also well known by now that after annealing the HPT processed sample, defects, introduced via SPD, anneal out partially and cracks shrink to pores, resulting in a smaller lattice parameter and higher density. It is important to note that neither the lattice parameter nor the density reaches the original size, which indicates that not all effects introduced via SPD are lost after annealing. After this measurement induced annealing process the sample is stable. It was shown for several times that neither the lattice parameter nor the density show further changes within the error bar after repeating the heat treatment within the same temperature range. This also holds for all physical (thermoelectric) properties [55,65-68].

**3.3. Transport properties**

Temperature dependent electrical resistivities in a temperature range of 300 – 823 K are shown in Fig.3a for $DD_{0.59}Fe_{2.7}Co_{1.3}Sb_{11.8}Sn_{0.2}$ and $DD_{0.54}Fe_{2.7}Co_{1.3}Sb_{11.9}Ge_{0.1}$ and are compared with the resistivity data of undoped $DD_{0.60}Fe_{2.8}Co_{1.2}Sb_{12}$ [50]. Both doped samples have a higher electrical resistivity than the undoped skutterudite due to a lower hole concentration. After annealing, the electrical resistivity of $D_{0.59}Fe_{2.7}Co_{1.3}Sb_{11.8}Sn_{0.2}$ is practically the same, but for $DD_{0.54}Fe_{2.7}Co_{1.3}Sb_{11.9}Ge_{0.1}$ an enhancement, increasing with increasing temperature was detected, with values now in the range of those of $D_{0.59}Fe_{2.7}Co_{1.3}Sb_{11.8}Sn_{0.2}$.

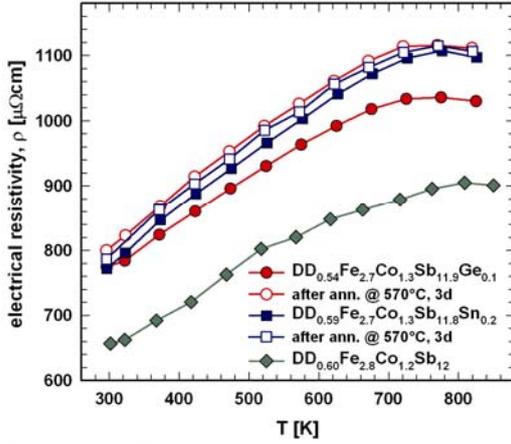 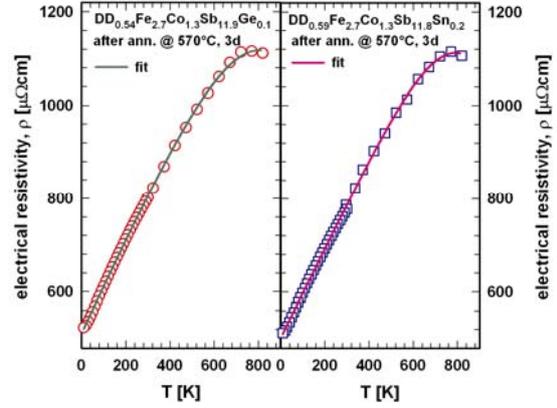

Fig. 3a. Temperature dependent electrical resistivity, ρ(T), above room temperature of $DD_{0.59}Fe_{2.7}Co_{1.3}Sb_{11.8}Sn_{0.2}$ and $DD_{0.54}Fe_{2.7}Co_{1.3}Sb_{11.9}Ge_{0.1}$ in comparison to $DD_{0.60}Fe_{2.8}Co_{1.2}Sb_{12}$ [50].

Fig. 3b. Temperature dependent electrical resistivity, ρ(T), with fits (see the text) of $DD_{0.59}Fe_{2.7}Co_{1.3}Sb_{11.8}Sn_{0.2}$ and $DD_{0.54}Fe_{2.7}Co_{1.3}Sb_{11.9}Ge_{0.1}$ in the temperature range of 4.2 – 823 K.

Figs. 3b display the temperature dependent electrical resistivities ρ(T) of $DD_{0.59}Fe_{2.7}Co_{1.3}Sb_{11.8}Sn_{0.2}$ and $DD_{0.54}Fe_{2.7}Co_{1.3}Sb_{11.9}Ge_{0.1}$ in the range of 4.2 – 823 K. Basically both ρ(T) curves show two different regimes: a metallic like at temperatures up to ≈ 300 K and a degenerated semiconducting like above room temperature. This rather complicated temperature dependence can be explained from a model [93], which combines the description of a simple metal via the Bloch-Grüneisen law with a temperature dependent charge carrier density. The electronic density of states is represented by a rectangular band with a narrow region of unoccupied states and separated by a gap from the conduction band. The charge carrier density n(T) is calculated using Fermi-Dirac statistics. Metallic conduction is possible at low temperatures with the narrow band above the Fermi energy ensuring unoccupied states for the scattered charge carriers. As soon as this part of the density of states becomes occupied, further carriers have to be promoted across the gap. Following these assumptions, ρ(T) can be calculated from

$$\rho(T) = \frac{\rho_0 n_0 + \rho_{ph}}{n(T)}. \qquad (2)$$

$$n(T) = (n_n(T)n_p(T))^{1/2} + n_0 \qquad (3)$$

where n(T) is the temperature dependent carrier concentration, $n_0$ is a residual temperature independent charge carrier density, $n_n$ is the density of electrons and $n_p$ is the density of holes.

The residual resistivity $\rho_0$ originates from the scattering of electrons on impurity atoms, crystal defects and grain boundaries and is independent of the temperature, whereas $\rho_{ph}(T)$ denotes the scattering resulting from lattice vibrations accounted for in terms of the Bloch-Grüneisen law,

$$\rho_{(ph)}(T) = \Re \left(\frac{T}{\theta_D}\right)^5 \int_0^{\frac{\theta_D}{T}} \frac{z^5}{(e^z-1)(1-e^z)} dz \qquad (4)$$

(with $z = \hbar\omega/k_B T$, $\omega$ is the phonon frequency, $\theta_D$ = Debye temperature, $\Re$ = temperature independent electron-phonon interaction constant, $\hbar$ is the reduced Planck's constant, $k_B$ is the Boltzmann constant

Least squares fits were applied to the experimental data in the range of 4.2-825 K using equations (2-4) and the results are shown as lines in Figs. 3b. The residual resistivities are $\rho_0$ = 504 µΩcm for $DD_{0.59}Fe_{2.7}Co_{1.3}Sb_{11.8}Sn_{0.2}$ and $\rho_0$ = 525 µΩcm for $DD_{0.54}Fe_{2.7}Co_{1.3}Sb_{11.9}Ge_{0.1}$, the respective Debye temperatures are $\theta_D$ = 121 K and $\theta_D$ = 129 K. A fit performed for the regime below 300 K only (not shown in Fig. 3) led, within the error bar, to the same results.

Fig. 4 shows the dependence of the Seebeck coefficient on temperature. The positive values indicate typical p-type conduction. Both doped samples, before and after annealing, display first a linear increase of the Seebeck coefficient with temperature gradually increasing, then developing a maximum at about 700 K followed by a slight decrease. For the Sn-doped sample (S$_{max}$= 186 µV/K at 721 K) there is almost no difference before and after annealing. The Ge-doped sample shows a much higher Seebeck coefficient after annealing (S$_{max}$= 182 µV/K at 671 K). This enhancement may be due to a slight change in the composition during the heat treatment.

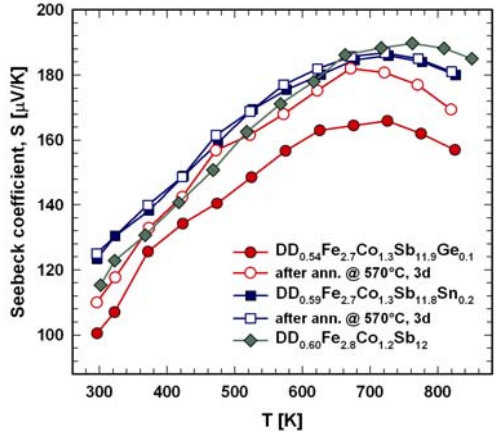 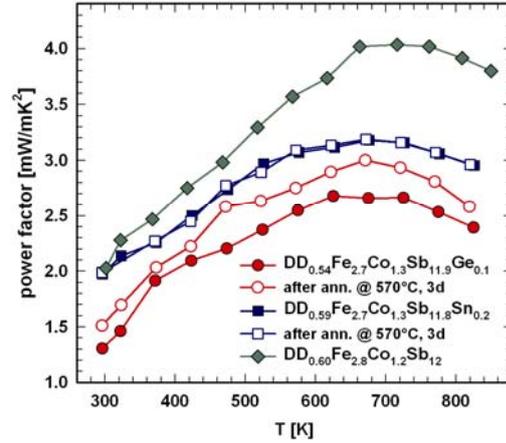

Fig. 4. Temperature dependent thermopower, S(T), of DD$_{0.59}$Fe$_{2.7}$Co$_{1.3}$Sb$_{11.8}$Sn$_{0.2}$ and DD$_{0.54}$Fe$_{2.7}$Co$_{1.3}$Sb$_{11.9}$Ge$_{0.1}$ in comparison to DD$_{0.60}$Fe$_{2.8}$Co$_{1.2}$Sb$_{12}$ [50].

Fig. 5. Temperature dependent power factor, S(T), of DD$_{0.59}$Fe$_{2.7}$Co$_{1.3}$Sb$_{11.8}$Sn$_{0.2}$ and DD$_{0.54}$Fe$_{2.7}$Co$_{1.3}$Sb$_{11.9}$Ge$_{0.1}$ in comparison to DD$_{0.60}$Fe$_{2.8}$Co$_{1.2}$Sb$_{12}$ [50].

The maximum of the Seebeck coefficient, S$_{max}$, can be used, as shown by Goldsmid and Sharp [94], to estimate the energy gap E$_g$ from

$$S_{max} = \frac{E_g}{2eT_{max}} \qquad (5)$$

where $e$ is the elementary charge and $T_{max}$ is the absolute temperature at which $S_{max}$ occurs. $E_g$ before and after annealing changes only marginally with $E_g$ = 243 ± 10 meV for the Ge-doped and $E_g$ = 265 ± 10 meV for the Sn-doped alloy. Both values are slightly lower as compared to DD$_{0.60}$Fe$_{2.8}$Co$_{1.2}$Sb$_{12}$ with $E_g$ = 290 ± 10 meV but are in the same range with the values gained from the resistivity fits.

The power factor (Fig. 5), for both doped samples is lower than for the undoped skutterudite DD$_{0.60}$Fe$_{2.8}$Co$_{1.2}$Sb$_{12}$ [50]. The same behavior was also found for the $Nd_{0.6}Fe_2Co_2Sb_{12-x}Ge_x$ series investigated by Zhang et al. [61]. But while in this series of skutterudites all power factors are below 2.2 mW/mK$^2$, the power factor of the annealed $DD_{0.54}Fe_{2.7}Co_{1.3}Sb_{11.9}Ge_{0.1}$ sample reaches 3.0 mW/mK$^2$, which is slightly higher than before annealing (2.7 at 672 K). For $DD_{0.59}Fe_{2.7}Co_{1.3}Sb_{11.8}Sn_{0.2}$ a high power factor of 3.2 mW/mK$^2$ at 672 K was reached, remaining unchanged for the annealed sample. The Seebeck coefficients of $DD_{0.59}Fe_{2.7}Co_{1.3}Sb_{11.8}Sn_{0.2}$ and $DD_{0.54}Fe_{2.7}Co_{1.3}Sb_{11.9}Ge_{0.1}$ are in the same range as those of the $Nd_{0.6}Fe_2Co_2Sb_{12-x}Ge_x$ skutterudites [61], however the latter have much higher electrical resistivities, explaining the lower power factors.

From the almost linearly varying S(T) one can extract the number n of charge carriers using Mott's formula

$$S = \frac{\pi^2 k_B^2 2m^*}{|e|\hbar^2 (3n\pi^2)^{2/3}} T, \tag{6}$$

(with m* = $m_e$ = 9.1094$^{-31}$ kg and |e| = 1.6022·10$^{-19}$A.s). n was calculated for the room temperature region, revealing average values of n ~ 1.8 × 10$^{21}$ cm$^{-3}$ and n ~ 1.5 × 10$^{21}$ cm$^{-3}$ for the Sn- and the Ge-doped skutterudite, respectively. Both data are only slightly lower than the values determined from Hall effect measurements shown in Fig. 6a. The temperature dependent charge carrier densities are increasing with increasing temperature. For 300 K n = 2.7 × 10$^{21}$ cm$^{-3}$ for $DD_{0.59}Fe_{2.7}Co_{1.3}Sb_{11.8}Sn_{0.2}$ and n = 2.4 × 10$^{21}$ cm$^{-3}$ for $DD_{0.54}Fe_{2.7}Co_{1.3}Sb_{11.9}Ge_{0.1}$. The Sn-doped sample has a slightly higher charge carrier density in the whole temperature regime measured than the Ge-doped sample. The distinct variation of the charge carrier concentration corroborates the earlier assumption of a gap in the density of states located slightly above $E_F$. Comparing n with the measured thermopower at room temperature, one can see an inverse dependence i.e. lower n yields higher S and vice versa.

The Hall mobility, $\mu = 1/\rho ne$, is plotted versus temperature in Fig. 6b. The mobility of both skutterudites is decreasing with increasing temperature and quite similar for both samples. In the double-logarithmic plot (insert in Fig. 6b) it can be seen that only in a very small temperature region the mobility exhibits a $T^{-1/2}$ dependence attributed to alloy scattering. The dependence on $T^{-2/3}$, typical of acoustic phonon scattering [95], cannot be observed in this plot.

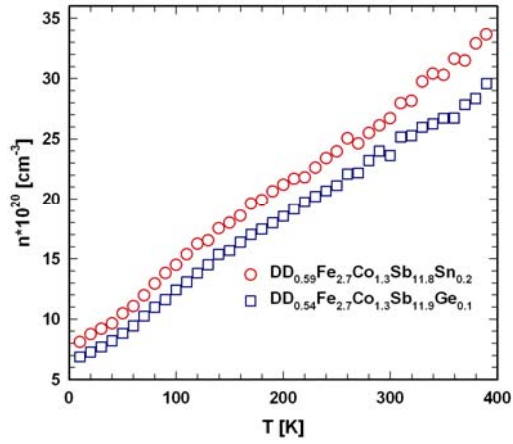

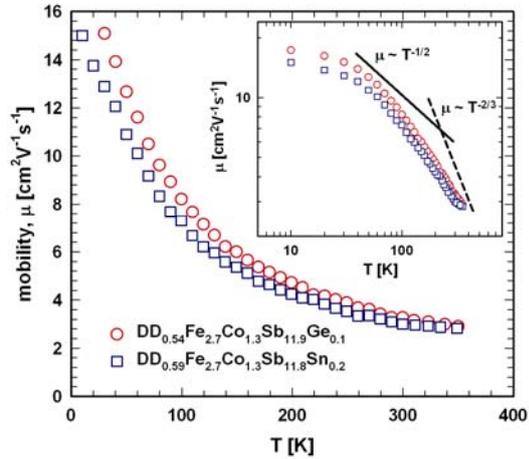

Fig. 6a. Temperature dependent number of charge carriers, n, of annealed $DD_{0.59}Fe_{2.7}Co_{1.3}Sb_{11.8}Sn_{0.2}$ and $DD_{0.54}Fe_{2.7}Co_{1.3}Sb_{11.9}Ge_{0.1}$.

Fig. 6b. Temperature dependent mobility, $\mu$, of annealed $DD_{0.59}Fe_{2.7}Co_{1.3}Sb_{11.8}Sn_{0.2}$ and $DD_{0.54}Fe_{2.7}Co_{1.3}Sb_{11.9}Ge_{0.1}$. Insert: double-logarithmic plot of $\mu(T)$ and two different temperature dependencies: $T^{-1/2}$ (solid line), $T^{-2/3}$ (dashed line).

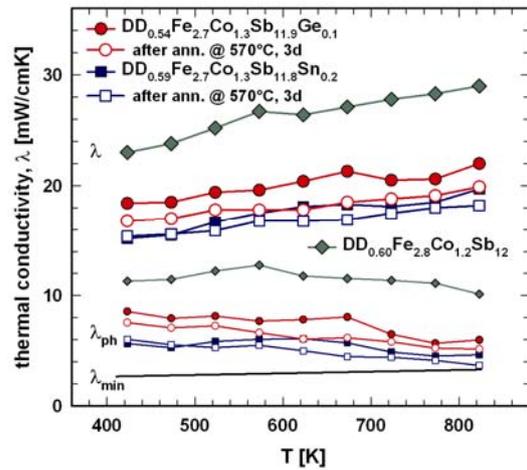

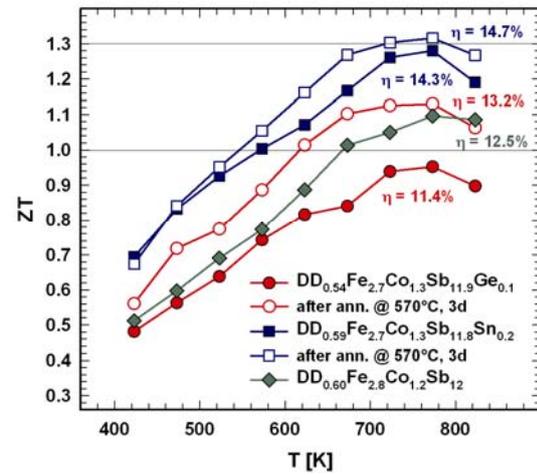

Fig. 7. Temperature dependent thermal, $\lambda(T)$, and lattice thermal conductivity, $\lambda_{ph}(T)$, of $DD_{0.59}Fe_{2.7}Co_{1.3}Sb_{11.8}Sn_{0.2}$ and $DD_{0.54}Fe_{2.7}Co_{1.3}Sb_{11.9}Ge_{0.1}$ in comparison to $DD_{0.60}Fe_{2.8}Co_{1.2}Sb_{12}$ [50].

Fig. 8. Temperature dependent figure of merit, ZT, of $DD_{0.59}Fe_{2.7}Co_{1.3}Sb_{11.8}Sn_{0.2}$ and $DD_{0.54}Fe_{2.7}Co_{1.3}Sb_{11.9}Ge_{0.1}$ in comparison to $DD_{0.60}Fe_{2.8}Co_{1.2}Sb_{12}$ [50].

Thermal conductivity (Fig. 7) is slightly increasing with increasing temperature in the whole temperature range measured. However, thermal conductivity for both doped samples is significantly reduced, by almost 35%, when compared with the undoped skutterudite. Obviously the Sb$_4$ rings, responsible for heat conduction, are disturbed by the induced Ge/Sn disorder [70,96] as also reported for Nd$_{0.6}$Fe$_2$Co$_2$Sb$_{12-x}$Ge$_x$ [61]. The lattice thermal conductivity, $\lambda_{ph}$, was obtained by subtracting the electronic contribution $\lambda_e$, from the total thermal conductivity $\lambda$ assuming the validity of the Wiedemann-Franz law, $\lambda_e \approx L_0 T/\rho$ with the Lorenz number $L_0 = 2.0 \times 10^{-8}$ W$\Omega$K$^{-2}$ as estimated by Chaput et al. [97].

As can be seen in the lower part of Fig. 4, $\lambda_{ph}$ is slightly decreasing with increasing temperature. Doping enhances the scattering of phonons on impurities and reduces the lattice thermal conductivity. For DD$_{0.59}$Fe$_{2.7}$Co$_{1.3}$Sb$_{11.8}$Sn$_{0.2}$ $\lambda_{ph}$ = 3 mW/cmK (at 425 K) for the HP sample and even lower with $\lambda_{ph}$ = 2.4 mW/cmK (at 425 K) after annealing the HP sample. $\lambda_{ph}$ for the Ge-doped sample is only slightly higher. Above 700 K the lattice thermal conductivity of both skutterudites before and after annealing gets close to $\lambda_{min}$. $\lambda_{min}$ is a theoretical value, which is close to the thermal conductivity of a glass and was estimated following the formula derived by Cahill and Pohl [98]

$$\lambda_{min} = \left(\frac{3n}{4\pi}\right)^{1/3} \frac{k_B^2 T^2}{\hbar \theta_D} \int_0^{\theta_D/T} \frac{x^3 e^x}{(e^x - 1)^2} dx \qquad (7)$$

where n = N/V is the number of atoms per unit volume, k$_B$ is the Boltzmann constant, $\hbar$ is the reduced Plancks constant, $\theta_D$ is the Debye temperature and x = $\hbar\omega/k_B T$.

Although the Seebeck coefficient was about the same for the doped and the undoped samples and the electrical resistivity was higher for the doped compounds, due to low thermal conductivities all ZT values of the doped skutterudites, except for one, are higher than that of the reference sample (Fig.8). ZT = 1.3 at 780 K is the highest value, gained for $DD_{0.59}Fe_{2.7}Co_{1.3}Sb_{11.8}Sn_{0.2}$ after annealing, but the value before annealing, considering error bars, is practically the same. This is an outstanding high ZT for a p-type skutterudite to the knowledge of the authors, the highest at the time.

The thermal-electric conversion efficiency with η = 12.5% for the reference sample $DD_{0.60}Fe_{2.8}Co_{1.2}Sb_{12}$ could be topped with η = 14.3% for $D_{0.59}Fe_{2.7}Co_{1.3}Sb_{11.8}Sn_{0.2}$ and is even higher with η = 14.7% for the annealed sample, which is an outstanding high value (all η-values calculated for 300-800 K).

The Ge-doped skutterudite (after annealing) with ZT = 1.1 in the temperature range of about 650-780 K and η = 13.2%, after all also has also a very good TE performance and ZT is in the range of $Nd_{0.54}Fe_{1.95}Co_{2.05}Sb_{11.99}Ge_{0.08}$ of Zhang et al. [61].

**3.3. Changes in transport properties after SPD via HPT**

TEM investigations were conducted exemplarily on Ge-doped samples after HPT (Fig. 9a) and after HPT with subsequent heat treatment (HT2) (Fig. 9b). Standard TEM specimen preparation by polishing and ion milling failed due to the micro cracks introduced by HPT. Instead, the samples were crushed into powders and dispersed on a holey carbon TEM grid. Both samples revealed grain sizes in the range of 15-300 nm with an average value of about 50 nm. Dislocations could not be imaged by diffraction contrast since the small grains, being significantly smaller than the SAED aperture, could not be tilted in defined diffraction conditions. Therefore, dislocation densities could not be determined. The superimposed energy-filtered RGB TEM images (Figs. 9a,b) revealed secondary phases (green colored) 20-50 nm in size dispersed in the matrix (purple colored). The secondary phases could be identified as $NdSb_2$ by EDX, they also showed a 3-6 times higher Ge content than the matrix.

$Fe_3Ge_2Sb$ secondary phases were found by XRD but could not be identified with the selected imaging conditions.

For both samples all transport properties were measured twice: electrical resistivity and in parallel thermopower were determined with increasing and decreasing temperature, thermal conductivity was measured with two runs. ~~In case of $DD_{0.54}Fe_{2.7}Co_{1.3}Sb_{11.9}Ge_{0.1}$ during the measurement of electrical resistivity and thermopower with decreasing temperature the sample broke, therefore data below 475 K are missing.~~

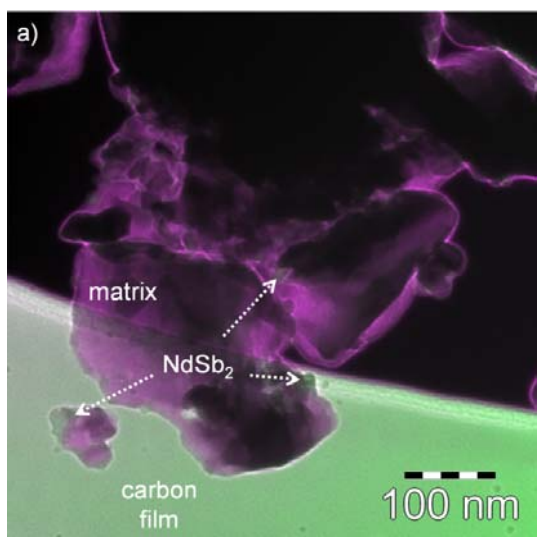 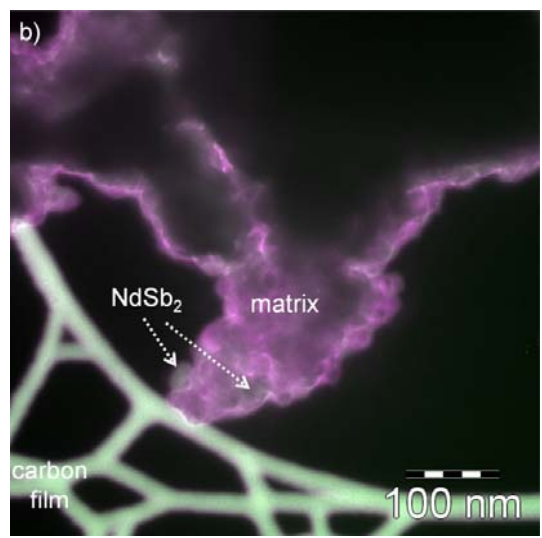

Fig. 9a. Superimposed energy-filtered (17, 31, and 17 eV) TEM images (RGB) of Ge-doped samples after HPT.

Fig. 9b. Superimposed energy-filtered (17, 31, and 17 eV) TEM images (RGB) of Ge-doped samples after HPT and heat treatment (HT2).

The SPD induced defects, i.e. smaller grains, an enhanced dislocation density and micro cracks affect not only the lattice parameter and density of a sample (see Figs. 2a and 2b) but also - to some extent - the transport properties. The changes after SPD i.e. introduced defects, smaller grains, an enhanced dislocation density and micro cracks affect not only the lattice parameter and density of a sample (see Figs. 2a and 2b) but also the transport properties to some extent. As can be seen in Fig. 11, severe plastic deformation does not much affect the Seebeck coefficient, S, exhibiting practically the same values as before HPT-processing within an error of 3% for the Sn-doped alloy. For the Ge-doped sample S is only slightly higher at T>700 K for in- and decreasing temperature.

However, a significant change takes place in the temperature dependent electrical resistivity curve (Fig. 10). Already at room temperature the electrical resistivity, ρ, is three to four times higher than before HPT. With increasing temperature ρ is increasing (more in case of the Ge-doped sample than in case of the Sn-doped sample) reaching a plateau like maximum and then decreases with values still higher than before HPT at 823 K. Such a behavior is already known from previous studies [55, 65-68] for filled and unfilled skutterudites [70,78] as well as for clathrates [92].

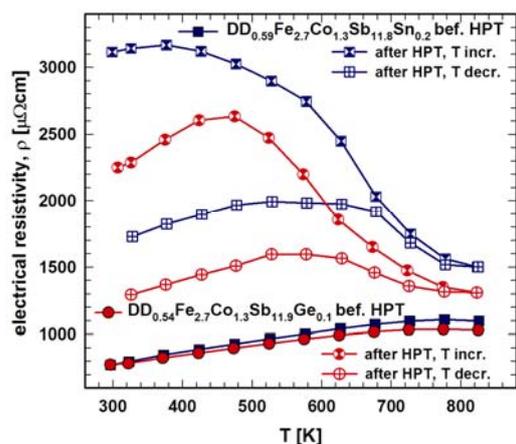
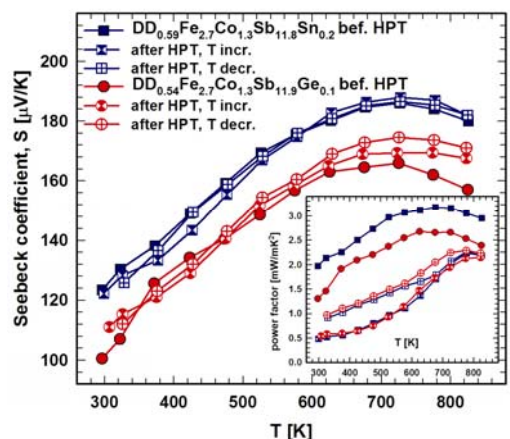

Fig. 10. Temperature dependent electrical resistivity, ρ(T), of D$_{0.59}$Fe$_{2.7}$Co$_{1.3}$Sb$_{11.8}$Sn$_{0.2}$ and DD$_{0.54}$Fe$_{2.7}$Co$_{1.3}$Sb$_{11.9}$Ge$_{0.1}$ before and after HPT processing.

Fig. 11. Temperature dependent Seebeck coefficient, S(T), of D$_{0.59}$Fe$_{2.7}$Co$_{1.3}$Sb$_{11.8}$Sn$_{0.2}$ and DD$_{0.54}$Fe$_{2.7}$Co$_{1.3}$Sb$_{11.9}$Ge$_{0.1}$ before and after HPT processing. Insert: temperature dependent power factor.

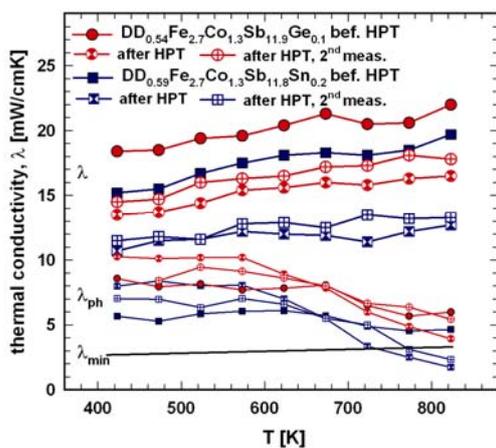
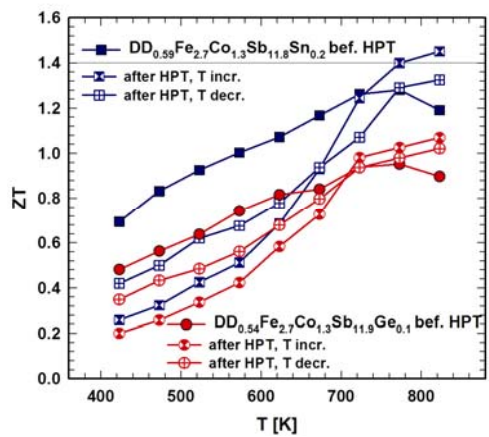

Fig. 12. Temperature dependent thermal

Fig. 13. Temperature dependent figure of merit, ZT,

conductivity, λ(T), and lattice thermal conductivity, λ$_{ph}$(T), of D$_{0.59}$Fe$_{2.7}$Co$_{1.3}$Sb$_{11.8}$Sn$_{0.2}$ and DD$_{0.54}$Fe$_{2.7}$Co$_{1.3}$Sb$_{11.9}$Ge$_{0.1}$ before and after HPT processing.

of D$_{0.59}$Fe$_{2.7}$Co$_{1.3}$Sb$_{11.8}$Sn$_{0.2}$ and DD$_{0.54}$Fe$_{2.7}$Co$_{1.3}$Sb$_{11.9}$Ge$_{0.1}$ before and after HPT processing.

After measurement induced heating, when most defects are annealed out, and cracks are fused to small pores (see also Ref. 99), the electrical resistivity decreases with decreasing temperature. It should be noted that between 675 and 855 K there is almost no difference in the electrical resistivity for the increasing and decreasing temperatures. Of course this high electrical resistivity contributes negatively to the power factor (insert in Fig. 11). Thermal conductivity (Fig. 12) is positively influenced by the introduced defects and accordingly is 20-30% lower after HPT in the whole temperature range, not changing much after measurement induced heating. Due to the high electrical resistivity in the lower temperature regime, the phonon part of the thermal conductivity, λ$_{ph}$, is slightly higher than before HPT, whereas above 723 K there is no difference, considering a 5% error. In case of the Sn-doped sample λ$_{ph}$ becomes even lower and is in the range of λ$_{min}$. This shows that a limit has been reached.

ZT of the SPD samples (Fig. 13) benefits from the lower thermal conductivity but suffers from the very high electrical resistivity. As the thermal conductivity of the doped skutterudites was already very low before HPT processing, the additional drop in thermal conductivity was not as spectacular as for many other filled skutterudites, nevertheless ZT of all HPT processed samples above ~ 700 K is higher than ZT of the original samples. ZT of DD$_{0.59}$Fe$_{2.7}$Co$_{1.3}$Sb$_{11.8}$Sn$_{0.2}$ surpasses 1.4 in the range of 773-823 K.

It should be added that after measurement induced heating the physical properties of HPT processed samples are stable as shown in earlier investigations of the authors [5-68]. Also in this work the data of a third and fourth measurement of electrical resistivity, Seebeck coefficient and thermal conductivity were within the error bar the same as those of the second one.

**Conclusions**

In order to further enhance ZT of p-type skutterudites, samples with the nominal composition $DD_{0.7}Fe_{2.7}Co_{1.3}Sb_{11.7}\{Ge,Sn\}_{0.3}$ were prepared and their structural and transport properties were studied. XRD combined with TEM investigations on $DD_{0.7}Fe_{2.7}Co_{1.3}Sb_{11.7}Ge_{0.3}$ samples yielded grain and crystallite sizes of 1-2 μm and about 340 nm after hot pressing and of about 300 nm and 100 nm after HPT processing respectively. Secondary phases, $Fe_3Ge_2Sb$ and $NdSb_2$, were found after HP, and although they annealed almost completely out after the first heat treatment, they were still present as traces after HPT processing, with a diameter of 20-50 nm and a 3-6 times higher Ge mole fraction than in the matrix. Transport measurements yielded a ZT~1.1 and a thermal-electric conversion efficiency (300-800 K) of η = 13.2% for the annealed sample with not much change after HPT processing.

The Sb/Sn doped sample was single-phase with the composition $DD_{0.59}Fe_{2.7}Co_{1.3}Sb_{11.8}Sn_{0.2}$. In comparison to the undoped skutterudite $DD_{0.60}Fe_{2.8}Co_{1.2}Sb_{12}$ the Sn-doped $DD_{0.59}Fe_{2.7}Co_{1.3}Sb_{11.8}Sn_{0.2}$ had a higher electrical resistivity, about the same Seebeck coefficient but a much lower thermal and lattice thermal conductivity, leading to a higher ZT~1.3, which to our knowledge is the highest ZT for a p-type skutterudite. Annealing at 570 K for 3 days proved the stability of the sample. Although there is hardly a change in ZT after annealing the sample, the thermal-electric conversion efficiency (300-800 K) with η = 14.3% is increased to η = 14.7%. After SPD, via HPT, because of additionally introduced defects, the lattice parameter increased, the density decreased and so did the crystallite size, as shown by XRD investigations. As a consequence of the introduced defects and/or cracks the electrical resistivity was enhanced, compensated by a much lower thermal conductivity. The net effect led to an even higher figure of merit (ZT = 1.45). After measurement induced heating most defects annealed out, leading to a lower electrical resistivity accompanied by a slightly higher thermal conductivity and because of a stable Seebeck coefficient to a lower ZT (ZT$_{stable}$ = 1.32).

**Acknowledgements:**

Research was supported by the "Christian Doppler Laboratory for Thermoelectricity" and FWF P24380. The authors thank Prof. R.C. Mallik and his team of the Indian Institute of Science, Dept. of Physics in Bangalore, for S- and $\rho$-measurements.

**Figure Captions**

| Fig. 1 | X-ray intensity pattern of D$_{0.59}$Fe$_{2.7}$Co$_{1.3}$Sb$_{11.8}$Sn$_{0.2}$ after HP, after HP + HT1, after HPT and after annealing the HPT sample (HPT + HT2). Insert: (240) peak after HP, after HPT and after annealing the HPT sample. |
|---|---|
| Fig. 2a | A: lattice parameter, a, B: relative density, d$_{rel}$, C: average crystallite size, s$_c$, D: grain size, s$_g$, of D$_{0.59}$Fe$_{2.7}$Co$_{1.3}$Sb$_{11.8}$Sn$_{0.2}$ as powder, HP, after HPT and after respective heat treatments. |
| Fig. 2b | A: lattice parameter, a, B: relative density, d$_{rel}$, C: average crystallite size, s$_c$, D: grain size, s$_g$, of D$_{0.54}$Fe$_{2.7}$Co$_{1.3}$Sb$_{11.9}$Ge$_{0.1}$ as powder, HP, after HPT and after respective heat treatments. |
| Fig. 3a | Temperature dependent electrical resistivity, $\rho(T)$, above room temperature of DD$_{0.59}$Fe$_{2.7}$Co$_{1.3}$Sb$_{11.8}$Sn$_{0.2}$ and DD$_{0.54}$Fe$_{2.7}$Co$_{1.3}$Sb$_{11.9}$Ge$_{0.1}$ in comparison to DD$_{0.60}$Fe$_{2.8}$Co$_{1.2}$Sb$_{12}$ [50]. |
| Fig. 3b | Temperature dependent electrical resistivity, $\rho(T)$, with fits (see the text) of DD$_{0.59}$Fe$_{2.7}$Co$_{1.3}$Sb$_{11.8}$Sn$_{0.2}$ and DD$_{0.54}$Fe$_{2.7}$Co$_{1.3}$Sb$_{11.9}$Ge$_{0.1}$ in the temperature range of 4.2 – 823 K. |
| Fig. 4 | Temperature dependent thermopower, S(T), of DD$_{0.59}$Fe$_{2.7}$Co$_{1.3}$Sb$_{11.8}$Sn$_{0.2}$ and DD$_{0.54}$Fe$_{2.7}$Co$_{1.3}$Sb$_{11.9}$Ge$_{0.1}$ in comparison to DD$_{0.60}$Fe$_{2.8}$Co$_{1.2}$Sb$_{12}$ [50]. |
| Fig. 5 | Temperature dependent power factor, S(T), of DD$_{0.59}$Fe$_{2.7}$Co$_{1.3}$Sb$_{11.8}$Sn$_{0.2}$ and DD$_{0.54}$Fe$_{2.7}$Co$_{1.3}$Sb$_{11.9}$Ge$_{0.1}$ in comparison to DD$_{0.60}$Fe$_{2.8}$Co$_{1.2}$Sb$_{12}$ [50]. |
| Fig. 6a | Temperature dependent number of charge carriers, n, of annealed DD$_{0.59}$Fe$_{2.7}$Co$_{1.3}$Sb$_{11.8}$Sn$_{0.2}$ and DD$_{0.54}$Fe$_{2.7}$Co$_{1.3}$Sb$_{11.9}$Ge$_{0.1}$. |
| Fig. 6b | Temperature dependent mobility, $\mu$, of annealed DD$_{0.59}$Fe$_{2.7}$Co$_{1.3}$Sb$_{11.8}$Sn$_{0.2}$ and DD$_{0.54}$Fe$_{2.7}$Co$_{1.3}$Sb$_{11.9}$Ge$_{0.1}$. Insert: double-logarithmic plot of $\mu(T)$ and two different temperature dependencies: T$^{-1/2}$ (solid line), T$^{-2/3}$ (dashed line). |

| Fig. 7 | Temperature dependent thermal, $\lambda(T)$, and lattice thermal conductivity, $\lambda_{ph}(T)$, of DD$_{0.59}$Fe$_{2.7}$Co$_{1.3}$Sb$_{11.8}$Sn$_{0.2}$ and DD$_{0.54}$Fe$_{2.7}$Co$_{1.3}$Sb$_{11.9}$Ge$_{0.1}$ in comparison to DD$_{0.60}$Fe$_{2.8}$Co$_{1.2}$Sb$_{12}$ [50]. |
|---|---|
| Fig. 8 | Temperature dependent figure of merit, ZT, of DD$_{0.59}$Fe$_{2.7}$Co$_{1.3}$Sb$_{11.8}$Sn$_{0.2}$ and DD$_{0.54}$Fe$_{2.7}$Co$_{1.3}$Sb$_{11.9}$Ge$_{0.1}$ in comparison to DD$_{0.60}$Fe$_{2.8}$Co$_{1.2}$Sb$_{12}$ [50]. |
| Fig. 9a | Superimposed energy-filtered (17, 31, and 17 eV) TEM images (RGB) of Ge-doped samples after HPT. |
| Fig. 9b | Superimposed energy-filtered (17, 31, and 17 eV) TEM images (RGB) of Ge-doped samples after HPT and heat treatment (HT2). |
| Fig. 10 | Temperature dependent electrical resistivity, $\rho(T)$, of D$_{0.59}$Fe$_{2.7}$Co$_{1.3}$Sb$_{11.8}$Sn$_{0.2}$ and DD$_{0.54}$Fe$_{2.7}$Co$_{1.3}$Sb$_{11.9}$Ge$_{0.1}$ before and after HPT processing. |
| Fig. 11 | Temperature dependent Seebeck coefficient, S(T), of D$_{0.59}$Fe$_{2.7}$Co$_{1.3}$Sb$_{11.8}$Sn$_{0.2}$ and DD$_{0.54}$Fe$_{2.7}$Co$_{1.3}$Sb$_{11.9}$Ge$_{0.1}$ before and after HPT processing. Insert: temperature dependent power factor. |
| Fig. 12 | Temperature dependent thermal conductivity, $\lambda(T)$, and lattice thermal conductivity, $\lambda_{ph}(T)$, of D$_{0.59}$Fe$_{2.7}$Co$_{1.3}$Sb$_{11.8}$Sn$_{0.2}$ and DD$_{0.54}$Fe$_{2.7}$Co$_{1.3}$Sb$_{11.9}$Ge$_{0.1}$ before and after HPT processing. |
| Fig. 13 | Temperature dependent figure of merit, ZT, of D$_{0.59}$Fe$_{2.7}$Co$_{1.3}$Sb$_{11.8}$Sn$_{0.2}$ and DD$_{0.54}$Fe$_{2.7}$Co$_{1.3}$Sb$_{11.9}$Ge$_{0.1}$ before and after HPT processing. |